\documentclass{article}

\usepackage{PRIMEarxiv}

\usepackage[utf8]{inputenc} % allow utf-8 input
\usepackage[T1]{fontenc}    % use 8-bit T1 fonts
\usepackage{hyperref}       % hyperlinks
\usepackage{url}            % simple URL typesetting
\usepackage{booktabs}       % professional-quality tables
\usepackage{amsfonts}       % blackboard math symbols
\usepackage{amsmath}
\usepackage{bm}
\usepackage{threeparttable}
\usepackage{nicefrac}       % compact symbols for 1/2, etc.
\usepackage{microtype}      % microtypography
\usepackage{lipsum}
\usepackage{fancyhdr}       % header
\usepackage{graphicx}       % graphics
\graphicspath{{media/}}     % organize your images and other figures under media/ folder

\title{Topological circuit of a versatile non-Hermitian quantum system
\thanks{\textit{\underline{Citation}}: 
\textbf{Galeano, DA., Zhang, XX. \& Mahecha, J. Topological circuit of a versatile non-Hermitian quantum system. Sci. China Phys. Mech. Astron. 65, 217211 (2022). https://doi.org/10.1007/s11433-021-1783-3}} 
}

\author{
  David-Andres Galeano \\
  Instituto de Física,
  Universidad de Antioquia,
  050010, Medell\'{\i}n, Colombia.\\
Gerencia Comercial T \& D Energía, 
V.P. T \& D Energía, EPM.
050015, Medell\'{\i}n, Colombia.\\
  \texttt{dandres.galeano@udea.edu.co} \\
   \And
 Xiao-Xiao Zhang \\
 Department of Physics and Astronomy \& Stewart Blusson Quantum Matter Institute,\\ University of British Columbia, Vancouver, British Columbia, V6T 1Z4 Canada.\\
  RIKEN Center for Emergent Matter Science (CEMS),
  Wako, Saitama 351-0198, Japan\\
    \texttt{xiaoxiao.zhang@riken.jp} \\
      \And
Jorge Mahecha\\
   Instituto de Física,
  Universidad de Antioquia,
  050010, Medell\'{\i}n, Colombia.\\
Department of Chemistry, University of British Columbia, Vancouver, British Columbia, V6T 1Z4, Canada.
  }
\begin{document}
\maketitle

\begin{abstract}
We propose an RLC (Resistors, Inductors and Capacitors) electrical circuit to theoretically analyze and fully simulate a new type of non-Hermitian Su-Schrieffer-Heeger (SSH) model with complex hoppings. We formulate its construction and investigate its properties by taking advantage of the circuit's versatility. Rich physical properties can be identified in this system from the normal modes of oscillation of the RLC circuit, including the highly tunable bulk-edge correspondence between topological winding numbers and edge states and the non-Hermitian skin phenomenon originating from a novel complex energy plane topology. The present study is able to show the wide and appealing topological physics inherent to electric circuits, which is readily generalizable to a plenty of both Hermitian and non-Hermitian nontrivial systems.
\end{abstract}

% keywords can be removed

\section{Introduction}\label{sec:intro}

\begin{figure}[h!]
\centering
\includegraphics[scale=0.7]{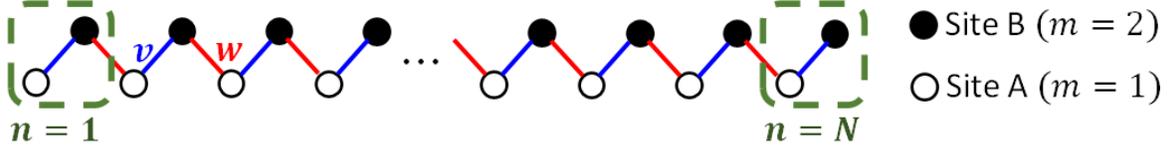}
\caption{SSH model for $N$ cells and hoppings between sites $A$ and $B$. Blue lines: intracell hoppings ($v$), and red lines: intercell hoppings ($w$).}
\label{fig:SSH_Hopping}
\end{figure}

The Hermitian Su-Schrieffer-Heeger (SSH) model \cite{SSH} describes fermionic spinless hopping in a one-dimensional (1D) lattice, as shown in Figure~\ref{fig:SSH_Hopping}. Non-Hermitian SSH models have been recently developed~\cite{Lee2019AnatomySystems,Yao2018EdgeSystems}, and several experiments have been performed with non-Hermitian SSH chains, such as photonic lattices \cite{Zeuner2015ObservationSystem, Weimann2017TopologicallyCrystals} or coupled dielectric microwave resonators \cite{Schomerus2013TopologicallyLattices,Poli2015SelectiveChain}, indicating the existence of topological zero modes and the presence of chiral symmetry.

Previous studies have shown that RLC circuits represent quantum mechanical phenomena~\cite{Alicata2018QuantumCircuits}, for example, topological phase transition for a system that exhibits non-Hermitian skin effects and Anderson localization behavior in a nonreciprocal Aubry-Andre model~\cite{Jiang2019InterplayLattices}, higher order topological Anderson insulators \cite{Zhang2021ExperimentalInsulators}, Chern insulators~\cite{Ezawa2019ElectricInsulators}, non-Hermitian Dirac and Weyl Hamiltonians~\cite{Luo2018TopologicalLattice, Zhang2020Non-HermitianCircuits}. Also, experiments with topoelectric circuits have been proposed to represent four-dimensional hexadecapole topological insulators \cite{Zhang2020Topolectrical-circuitInsulator}. Our main interest is to analyze underlying topological properties for non-Hermitian SSH chains using the correspondence with an RLC circuit chain and provide a representation of the associated non-conservative quantum behavior. Electromagnetic oscillation modes for RLC circuit chains have been extensively studied in electrical engineering~\cite{Parker1969NormalNetworks}. These modes depend exclusively on circuit component topologies and values, where the set of natural circuit frequencies correspond to the admittance function poles~\cite{Kuester2020TheoryLines}.

In the current study, the proposed RLC circuit corresponds to a new kind of non-Hermitian SSH Hamiltonian, i.e., $\hat{H}\neq \hat{H}^\dagger$, while preserving symmetries~\cite{Bergholtz2019ExceptionalSystems}. Non-Hermiticity changes some symmetries due to the distinction between complex conjugation and matrix transposition, which are equivalent in Hermitian Hamiltonians~\cite{Kawabata2019SymmetryPhysics}. Indeed, both Hermitian and non-Hermitian systems exhibit time reversal symmetry if the $\hat{\mathcal{T}}$-operator satisfies $[\hat{H},\hat{\mathcal{T}}]=0$ \cite{Shen2012}. In momentum space
$\hat{\mathcal{T}}\hat{H}^*(\mathbf{k})\hat{\mathcal{T}}^{-1}=\hat{H}(-\mathbf{k})$\cite{Ryu2010, Kaufmann2015}.
Particle-hole symmetry in a non-Hermitian Hamiltonian is related to charge conjugation that mixes fermion creation and annihilation operators and is defined by a unitary operator $\hat{\mathcal{P}}$. Specifically, particle-hole symmetry can be expressed in band theory as 
$\hat{\mathcal{P}} \hat{H}^T(\mathbf{k}) \hat{\mathcal{P}}^{-1} = -\hat{H}(-\mathbf{k})$\cite{Kawabata2019SymmetryPhysics}.
Chiral symmetry is usually a combination of time reversal and particle-hole symmetry~\cite{Kaufmann2015, Prodan2016} and defined in terms of an antiunitary operator $ \hat{\Gamma}$. Chiral symmetry in a non-Hermitian system implies that the eigenvalues appear in pairs, which is important for protected zero-energy edge modes~\cite{Ezawa2019Non-HermitianRealization}. Thus, chiral symmetry in non-Hermitian systems can be expressed as 
$\hat{\Gamma}\hat{H}^{\dagger}(\mathbf{k})\hat{\Gamma}^{-1}=-\hat{H}(\mathbf{k})$\cite{Kawabata2019SymmetryPhysics}.

Our proposed non-Hermitian circuit model (Section~\ref{sec:model}) has chiral symmetry. Disregarding interactions between electrons, the general SSH model dynamics can be described by the Hamiltonian~\cite{Asboth2015}
\begin{equation}\label{eq:SSH}
\hat{H}_{\textrm{SSH}} = v \sum_{n=1}^N \left( |n,B\rangle \langle n, A| + h.c    \right) + 
 w \sum_{n=1}^{N-1} \left( |n+1,A\rangle \langle n, B| + h.c    \right),
\end{equation}
where $|n,A\rangle$ and $|n,B\rangle$ ($n\in \{1,2,\cdots,N \}$) are lattice states, and the electron is in a unit cell $n$ at sublattice sites $A$ or $B$  (see Figure~\ref{fig:SSH_Hopping}). 
Hoppings $v$ and $w$ are real in Hermitian SSH models but can take complex values for non-Hermitian models. The usefulness of a model that involves complex hopping is given by the possibility of developing long-range hopping models in topological plasmonic chain \cite{Pocock2019Bulk-edgeChain}, interacting boson chains \cite{deLeseleuc2019ObservationAtoms}, polariton crystals \cite{Pickup2020SyntheticPhases}, among others. The Schr\"{o}dinger equation that defines the Hamiltonian (\ref{eq:SSH}) in the momentum-space is 
$\hat{H}(k)[a(k),b(k)]^\mathrm{T} = E(k)[a(k),b(k)]^\mathrm{T}$
with
\begin{equation}\label{eq:SH_Auto}
    \hat{H}(k) = \begin{pmatrix}
    0 & v+we^{-ik} \\
    v+we^{ik} & 0
    \end{pmatrix},
\end{equation}
$E(k)=\pm |v+we^{-ik}| = \pm \sqrt{v^2+w^2+2vw\cos{k}}$, and $[a(k),b(k)]^\mathrm{T}$ are the corresponding eigenvectors.

Lee~\cite{Lee2018TopolectricalCircuits} and Zhao~\cite{Zhao2018TopologicalCapacitors} represented a Hermitian SSH model using electrical circuits with LC components based on a tight-binding approximation. Zhao used charge and electric flux ($\varphi$) concepts to deduce the Euler-Lagrange equations. The circuital Hermitian model corresponding to the SSH proposed by Zhao becomes an LC circuit 
system where the bands are given by functions $\omega_\pm(k)$, 
\begin{equation}\label{eq:Zhao_Bands}
\frac{\omega_\pm^2(k)}{\omega_0^2} = \eta+\eta^{-1}\pm\sqrt{\eta^2 + 
    \eta^{-2}+2\cos{k}}, 
\end{equation}
where $\eta=\sqrt{L_1/L_2}$, $\omega_0 = 1/(C\sqrt{L_1L_2})$, $C$ is the capacitance connected to ground, and $L_i~i=\{1,2\}$ are the circuit inductors.

In contrast to previous Hermitian LC circuit models, we investigate the natural normal modes of a new type of non-Hermitian SSH model. It includes resistive components and furthermore, the non-Hermiticity appears with momentum dependence rather than being imaginary constants added to the two-band model. This allows new topological phases with the appearance of various winding number values ($\mu$)~\cite{Yin2018GeometricalSystems, Ghatak2019NewSystems, Gong2018TopologicalSystems}. In particular, the proposed circuital system maintains topological protection, i.e., the bulk gap is preserved, and the system remains robust against perturbations, hence the winding number cannot change continuously. 

In addition, this system is able to show a general non-Hermitian skin effect, in which many bulk states are pushed to the edges and the coexistence of skin effect and topological edge states becomes a remarkable feature \cite{Xu2021CoexistenceRealizations}. Thanks to the high tunability at the `site level' in circuits, which allows the full tomography of voltage/currents at circuit nodes and impedance between any nodes, the intricate information of the simulated eigenvectors becomes accessible \cite{Lee2018TopolectricalCircuits,Zhang2020Non-HermitianCircuits}. In the Appendix, we also provide the realistic circuit simulation, from which one can observe all the proposed important non-Hermitian topological features. We hence expect the present study to provide new insight into the non-Hermitian topological systems and to considerably widen the realistic investigations in the readily available circuit systems.

In the next section, we analyze our proposed model’s topological implications, emphasizing winding number, bulk-edge correspondence, and the skin effect. In particular, we did not focus on developments related to preservation or breaking bulk-boundary correspondence but rather showing that the natural RLC circuit model oscillation modes effectively represent non-Hermitian SSH system quantum effects and are consistent with theoretical expectations. 

%%%%%%%%%%%%%%%%%%%%%%%%%%%%%%%%%%%%%%%%

\section{Model}\label{sec:model}
Our main objective is to represent non-Hermitian system topological behavior through an RLC circuit’s natural oscillation modes. RLC circuit behavior is determined by its Laplacian~\cite{Lee2018TopolectricalCircuits}, hence this study establishes a bridge between electrical engineering and material science that will help to understand quantum system behaviors. Electrical circuits can be represented by a graph that links active and passive elements. Similarly, crystalline lattices have hoppings with special characteristics that can be emulated by circuital networks, known as topolectric circuits. This class of circuits allows the tight-binding approximation, which implies that position eigenstate dynamics $|x\rangle$ are restricted to lattice sites~\cite{Tong2016, Shen2012}.

\begin{figure}[h!]
\centering
\includegraphics[scale=0.5]{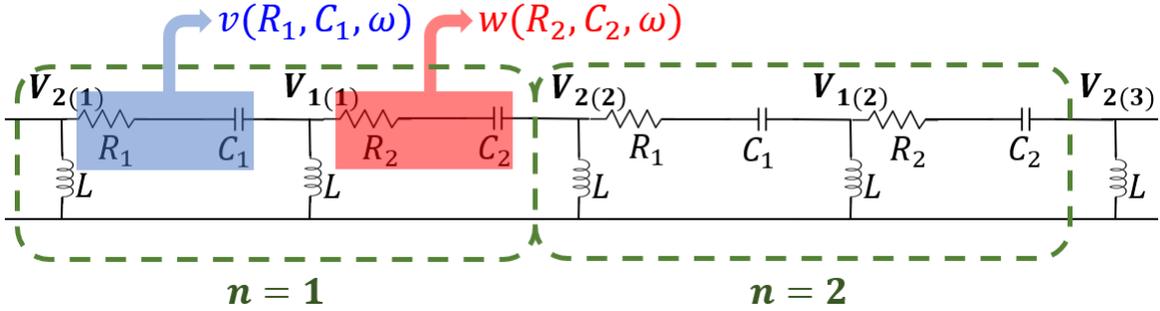}
\caption{Proposed non-Hermitian topolectric circuit. Each $n$-unit cell comprises a pair of resistors and capacitors with the same inductor L between each RC array. Circuit elements with subscript 1 (blue) represent intracell hoppings, and ones with subscript 2 (red) represent intercell hoppings. At each node, we use the notation $V_{m(n)}$, where $n$ is the cell, and $m$ is the site. }
\label{fig:Circuit_SSH}
\end{figure}

The SSH model includes intra-cell ($v$) and inter-cell ($w$) hopping, as discussed in Section~\ref{sec:intro} and shown in Figure~\ref{fig:SSH_Hopping}. If $v, w\in \mathbb{C}$, then Hamiltonian (\ref{eq:SSH}) becomes non-Hermitian~\cite{Li2014TopologicalModels}. 
Figure~\ref{fig:Circuit_SSH} shows the proposed RLC circuit that emulates a non-Hermitian SSH system. Kirchhoff's laws allow the RLC circuit to be expressed as a constructive block, emulating a tight-binding lattice. In particular, a hypothetical current $ I_j $ along  node $j$ of that fundamental unit is a component of vector $\bm{I}$, which is related to the corresponding vector of potential differences $\bm{V}$ by $\bm{I} = \mathcal{L}\bm{V}$, where matrix $\mathcal{L}$ is the circuit Laplacian and represents admittance between nodes; with $\mathcal{L}=D-C+W$, where $D$ is a diagonal matrix containing admittances for each node, $C$ is a matrix with zeroes on the diagonal and admittances between different nodes elsewhere, and $W$ is a matrix that relates the admittances that lead to the ground~\cite{Lee2018TopolectricalCircuits}. 

To explain the proposed model, we start with Kirchhoff's second law. We propose the following equations respectively for nodes $1(1)$ and $2(1)$ in Figure~\ref{fig:Circuit_SSH},
$
[V_{2(1)}-V_{1(1)}]Y_{RC_1} = [V_{1(1)}-V_{2(2)}]Y_{RC_2}+\frac{V_{1(1)}}{i\omega L}
$ and
$
[V_{1(1)}-V_{2(2)}]Y_{RC_2} = [V_{2(2)}-V_{1(2)}]Y_{RC_1}+\frac{V_{2(2)}}{i\omega L}  
$,
where $Y_{RC_j}=\frac{1}{Z_{RC_j}}=\frac{1}{R_j+(i\omega C_j)^{-1}}$ is the admittance for resistors and capacitors in series. Solving this expression requires appropriate periodic boundary conditions. We consider transformations $V_{j (n)}(t) = \frac{e^{-i\omega t}}{\sqrt{N}}\sum_k V_j'(k) e^{ikn}$, with $j=1,2$, and $k\in [0,2\pi]$ since the Fourier transform generates a map in the $k$ wave vector spectrum establishing a periodic band structure for the admittances in the circuit~\cite{Helbig2019BandNetworks}. Therefore, we obtain the following system 
\begin{equation}\label{eq:H}
    \frac{1}{i\omega}\mathcal{L}\;[V'_1(k),V'_2(k)]^\mathrm{T}=0
\end{equation}
with 
$\eta_1 = i\omega R_1 C_1 + 1,
\eta_2 = i\omega R_2 C_2 + 1$ and the Laplacian \begin{equation}\label{eq:L}
\begin{split}
     \mathcal{L} =  
    % \begin{pmatrix}
    % Y_{RC_1} + Y_{RC_2} + \frac{1}{i\omega L}  & -Y_{RC_1} - Y_{RC_2} e^{ik}\\
    % -Y_{RC_1} - Y_{RC_2} e^{-ik}  &  Y_{RC_1} + Y_{RC_2} + \frac{1}{i\omega L} \end{pmatrix}\\
    %  = & 
    i\omega \left(  \frac{C_1}{\eta_1}+\frac{C_2}{\eta_2}-\frac{1}{\omega^2L} \right)\mathrm{I} - 
     i\omega \left[ \left(\frac{C_1}{\eta_1} + \frac{C_2}{\eta_2}\cos{k}\right)\sigma_x + \left(\frac{C_2}{\eta_2}\sin{k}\right)\sigma_y  \right],
\end{split}
\end{equation}
where $\sigma_i$ are the Pauli matrices, with $i=\{x,y\}$. 
Comparing the non-Hermitian Laplacian (\ref{eq:L}) with Zhao and Lee \textit{et al.}'s Hermitian Laplacian~\cite{Zhao2018TopologicalCapacitors,Lee2018TopolectricalCircuits}, the Laplacian (\ref{eq:L}) coincides with the Hermitian Laplacian if $R_1=R_2 = 0$, i.e., there are no dissipative elements. 
We can also rewrite (\ref{eq:H}) 
 \begin{equation}\label{eq:H1}
    \Lambda(\omega)\;
    [V'_1(k),V'_2(k)]^\mathrm{T} = \mathcal{Y}(k)\;
   [V'_1(k),V'_2(k)]^\mathrm{T}, 
 \end{equation} 
 with the admittance matrix for the non-Hermitian circuit
\begin{equation}\label{eq:Hamiltonian}
    \mathcal{Y}(k)=\begin{pmatrix}
    0 & -\frac{C_1}{\eta_1}-\frac{C_2}{\eta_2}e^{-ik}\\
    -\frac{C_1}{\eta_1}-\frac{C_2}{\eta_2}e^{ik} & 0
    \end{pmatrix}
\end{equation}
and $\Lambda(\omega)= \left(\frac{1}{\omega^2 L} - \frac{C_1}{\eta_1}- \frac{C_2}{\eta_2} \right)$ is the sum of the admittances that converge at each node, which emulates the system bulk bands. Therefore, the bulk gap between bands is $2\Delta$, where $\Delta=\min_k\Lambda$.
The relationship between $\mathcal{Y}(k)$ and the SSH Hamiltonian (\ref{eq:SH_Auto}) is 
\begin{equation}\label{eq:hopping_vw}
\begin{split}
v&=-\frac{C_1}{\eta_1} = \sqrt{\frac{C_1^2}{1+C_1^2R_2^2\omega^2}}e^{-i\arctan{(C_1R_1\omega)}}\\
w&=-\frac{C_2}{\eta_2} = \sqrt{\frac{C_2^2}{1+C_2^2R_2^2\omega^2}}e^{-i\arctan{(C_2R_2\omega)}}.
\end{split}
\end{equation}
It is important that the proposed model has mutually independent intra ($v$) and inter ($w$) cell hopping, in contrast with previous models~\cite{Zhao2018TopologicalCapacitors, Lee2018TopolectricalCircuits}.

We rewrite (\ref{eq:H1}) as $\mathcal{Y}(k)|u_p(k)\rangle = \Lambda_p |u_p(k)\rangle$,
% \begin{equation}\label{eq:Dirac_Nota}
%     \mathcal{Y}(k)|u_p(k)\rangle = \Lambda_p |u_p(k)\rangle,
% \end{equation}
where index $p$ labels the bands. Consequently, this does not hold for arbitrary $\omega$, but only for normal modes. Therefore, we can calculate the bands from (\ref{eq:H1}) as
\begin{equation}\label{eq:omega}
\frac{\omega^2}{\omega_0^2} = \frac{1}{\gamma+\gamma^{-1}+ \sqrt{\gamma^2+\gamma^{-2}+2\cos{k}}},    
\end{equation}
where $\omega_0^2 = \frac{\sqrt{\eta_1\eta_2}}{L\sqrt{C_1C_2}}$, and $\gamma =\sqrt{\frac{C_1\eta_2}{C_2\eta_1}}$ are function of $\omega$. 
Although (\ref{eq:Zhao_Bands}) and (\ref{eq:omega}) have similar mathematical structures, $\omega_0$ is function of $\omega$ in (\ref{eq:omega}), whereas this does not hold for $(\ref{eq:Zhao_Bands})$. This dependence on $\omega$ makes a fundamental difference between Zhao’s Hermitian system~\cite{Zhao2018TopologicalCapacitors} and the non-Hermitian system proposed here, in addition to hopping properties.
Equation (\ref{eq:omega}) is a sixth degree polynomial for $\omega$ as a function of $k$. Two solutions correspond to bands in a complex plane that are independent of $k$, i.e., to bands at $\Lambda=\infty$ where
$\omega_1=\frac{i}{R_1C_1}, \omega_2=\frac{i}{R_2C_2}$.
Thus, the energy eventually stored by the inductors and capacitors is dissipated by the resistors. Therefore, we focused on analyzing the four remaining solutions for $\omega(k)$, called $\omega_3$, $\omega_4$, $\omega_5$, and $\omega_6$.
Normal oscillation modes should not be confused with the oscillations of decreasing complex exponential function from a source connected to the circuit. Rather, this term refers to the index of exponential functions that describe the natural source-free response when capacitors and/or inductors are charged~\cite{Kumar2009ElectricNetworks}.
\begin{figure}[h!]
\centering
\includegraphics[scale=0.7]{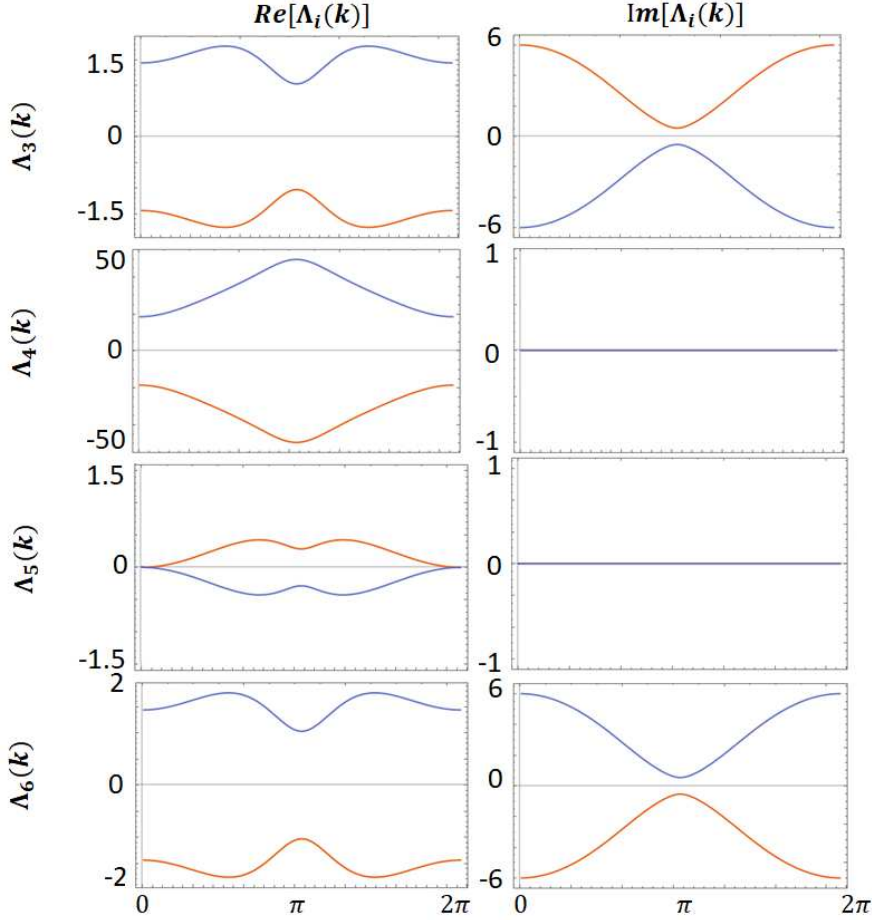}
\caption{Complex eigenvalues $\Lambda(k)$ from (\ref{eq:H1}) for each $\omega$ solution with parameters $R_1 = 3.2~\Omega$, $R_2=2.6~\Omega$, $C_1=3.4 ~F$, $C_2=3.1 ~F$, and $L=2.7~H$}
\label{fig:Lambda_P}
\end{figure}

\begin{figure}[h!]
\centering
\includegraphics[scale=0.99]{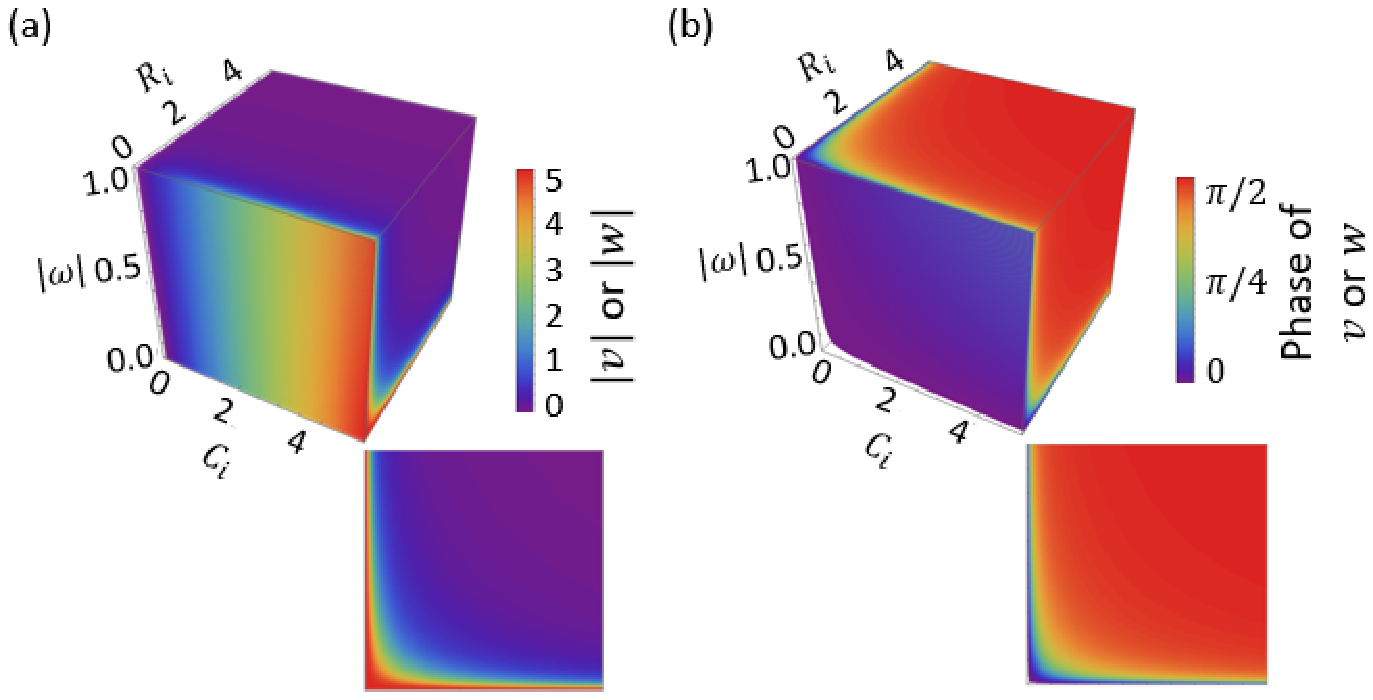}
\caption{Hopping characterization with respect to circuit parameters and complex frequency magnitude from (\ref{eq:hopping_vw}). (a) (top) Magnitude and (b) (top) hopping phase from $R_i$, $C_i$, and $|\omega|$. (a) (bottom) and (b) (bottom) magnitude and phase in the $R-|\omega|$ plane, respectively.}
\label{fig:Magn_Hoppings}
\end{figure}

\begin{figure}[h!]
\centering
\includegraphics[scale=0.99]{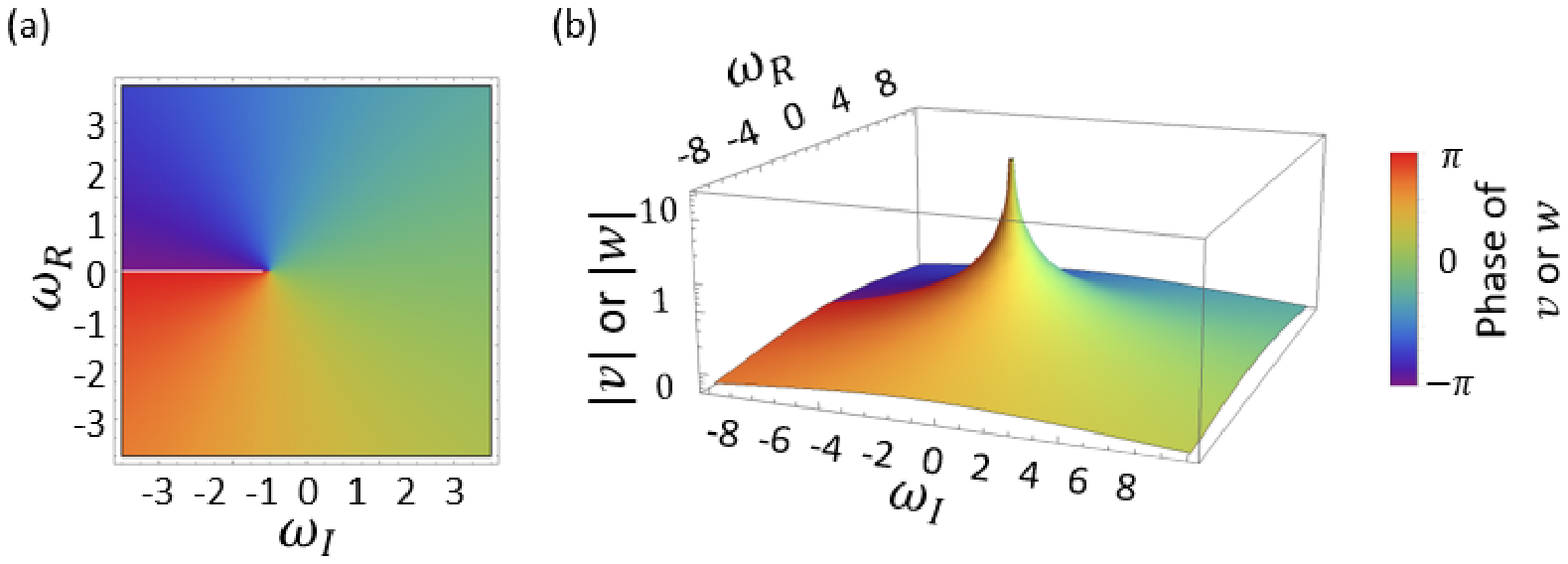}
\caption{Hopping (a) phase and (b) magnitude with respect to complex natural frequency $\omega=\omega_{R}+i\omega_{I}$}
\label{fig:Skin}
\end{figure}

A dissipative topolectric model can have real and in general complex admittance spectra for normal oscillation modes $\omega(k)$. From (\ref{eq:Hamiltonian}), our system allows for both spectral types for $\Lambda$. Figure~\ref{fig:Lambda_P} shows an example of the spectra corresponding to the aforementioned four normal modes. 
Figures~\ref{fig:Magn_Hoppings} and \ref{fig:Skin} show $v$ and $w$ behavior with a complex frequency. Although (\ref{eq:H1}) is true for natural oscillation modes, hopping exhibits a continuum in complex space and $\omega=\omega_R+i\omega_I$ due to freedom in choosing the circuit parameters.
Hopping $v$ and $w$ reduce as $\omega$ magnitude and $R_i$ increase. However, hopping quickly acquires a phase that tends to $\pi/2$. Figures~\ref{fig:Magn_Hoppings}(a) and \ref{fig:Magn_Hoppings}(b) show that hoppings tend to have similar behavior to a Hermitian system for $|\omega|$ close to zero and small $R_i$. 

The proposed model allows modulating hopping magnitude by choosing a natural oscillation frequency from (\ref{eq:omega}). Hopping magnitude reduces asymptotically to zero as $\omega$ magnitude increases (see Figure~\ref{fig:Skin}); tend to zero in the bulk in the hypothetical limit when $|\omega|\rightarrow\infty$~\cite{Song2019Non-HermitianSystems}; and rapidly approach  infinity when $\omega_{I}=\mathrm{Im}(\omega)\rightarrow -1$ and $\omega_{R}=\mathrm{Re}(\omega)\rightarrow 0$.

The proposed circuit model fits adequately to theoretical physical models. For example, Fu~\cite{Fu2020ExtendedCouplings} proposed a model that involves a binary waveguide array with alternating real and imaginary couplings, hence $v$ is imaginary $(R_1\gg C_1)$ and $w$ is real $(R_2\rightarrow 0)$ in our RLC circuit model. The physical model proposed by Pocock \textit{et al.}~\cite{Pocock2019Bulk-edgeChain} and subsequently Lieu~\cite{Lieu2018TopologicalModel} can also be adjusted to our proposed circuital model results. The non-Hermitian physical models for complex hopping that they analyzed, exhibited bulk-edge correspondence when the next nearest neighbor was zero, i.e., the Hamiltonian diagonal was zero, for any $v$ and $w$ magnitude and phase. Section~\ref{sec:Resultados} discusses this further along with other topological properties that emerge from the proposed circuit model that is aligned with physical phenomena.

%%%%%%%%%%%%%%%%%%%%%%%%%%%%%%%%%%

\section{Topological properties}\label{sec:Resultados}

We used the circuit model from Section~\ref{sec:model} to identify different topological properties of the RLC chain (Figure~\ref{fig:Circuit_SSH}) and the non-Hermitian SSH model. Chiral symmetry in a non-Hermitian SSH chain guarantees topological invariants since topologically protected edge modes are their own symmetric partners~\cite{Yao2018EdgeSystems}. We also discuss bulk-edge correspondence in this non-Hermitian model with periodic boundary conditions and discuss the relationship for these edge states with natural oscillation frequency magnitude.

\subsection{Non-Hermitian topological winding number and protected edge states}\label{sec:WN}
\begin{figure*}[t]
\centering
\includegraphics[scale=0.55]{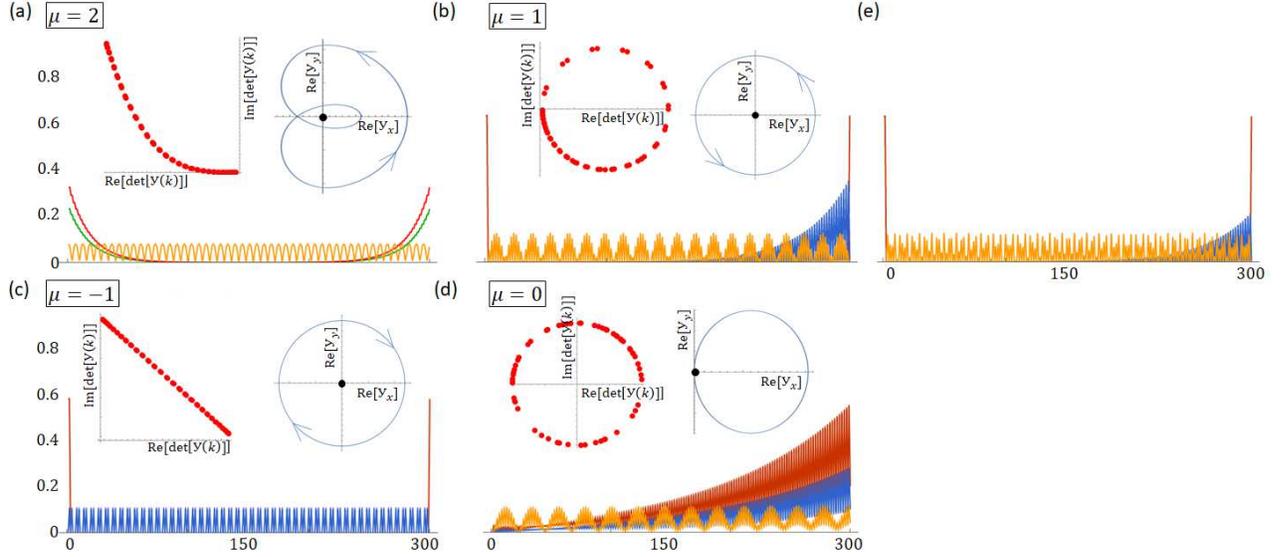}
\caption{Eigenvector magnitude for the parameter set from Table~\ref{tab:table2}, row 4. Each figure shows the trajectory of the admittance matrix determinant for skin effect existence (left inset) and the trajectory of the real part of $\mathcal{Y}(k)$ (indicating clockwise or counterclockwise) around the exception point for determining the winding number (right inset). We focus on a few low-energy states. (a)–(c) have topological edges states (the red and green isolated lines) and (b,d) show skin effect (the amplitude decaying modes along a certain direction). (e) shows the counterpart of (b) with small perturbations in a few circuit elements around the middle of the chain, where the topological edge states (red) remain stable while the Bloch states (yellow) and the skin effect (blue) are perturbed.}
\label{fig:SkinV2}
\end{figure*}

\begin{table}[h!]
\footnotesize
\begin{threeparttable}\caption{Circuit parameters with respective winding numbers for each solution $\omega_i(k)$.}\label{tab:table2}
\begin{tabular}{lccccccccr}
\toprule
$R_1~[\Omega]$ & $R_2~[\Omega]$ & $C_1~[F]$ & $C_2~[F]$ & $L~[H]$ & $\mu_3$ & $\mu_4$ & $\mu_5$ & $\mu_6$\\
\hline
1.34 & 0.17 & 0.95 & 0.45 & 0.81 & -1 & 1 & 1 & 0\\
0.03 & 0.14 & 1.50 & 0.26 & 0.57 & 0 & 0 & 2 & 0 \\
1.45 & 0.14  & 0.22 & 0.54 & 1.11 & 1 & 0 & 0 & 1 \\
0.05 & 1.41 & 0.03 & 1.34 & 1.17 & 2 & 1 & -1 & 0\\
\bottomrule
\end{tabular}
\end{threeparttable}
\end{table}
Since $\mathcal{Y}(k)$ can be written as a two-band non-Hermitian system,
$\mathcal{Y}(k)=\mathcal{Y}_x\sigma_x+\mathcal{Y}_y\sigma_y$ with $\mathcal{Y}_x=-\frac{C_1}{\eta_1}-\frac{C_2}{\eta_2}\cos{k},\mathcal{Y}_y=\frac{C_2}{\eta_2}\sin{k}$,
it follows that the proposed model has chiral symmetry $\sigma_z \mathcal{Y}^\dagger(k)\sigma_z=-\mathcal{Y}(k)$ and $\mathcal{Y}_x$ and $\mathcal{Y}_y$ close the loop in reciprocal space as $k$ varies from $0$ to $2\pi$,. 
Given the four $\omega_i$ families derived from (\ref{eq:omega}) and using Yin \textit{et al.}'s argument~\cite{Yin2018GeometricalSystems}, the winding number can be expressed as
\begin{equation}\label{eq:Winding_Number}
%\begin{split}
    \mu=%&\frac{1}{2\pi}\oint_c \frac{\mathcal{Y}_x d\mathcal{Y}_y - \mathcal{Y}_y d\mathcal{Y}_x}{\mathcal{Y}_x^2+\mathcal{Y}_y^2}\\=&
    \frac{1}{2\pi}\int_0^{2\pi} dk \frac{\mathcal{Y}_x \partial_k \mathcal{Y}_y - \mathcal{Y}_y \partial_k \mathcal{Y}_x}{\mathcal{Y}_x^2+\mathcal{Y}_y^2},
%\end{split}
\end{equation}
where there is a winding number for each family of normal oscillation frequencies. Therefore, there are four possible winding numbers for each $R_1$, $R_2$, $C_1$, $C_2$, and $L$ configuration. Table \ref{tab:table2} shows example winding numbers for a particular parameter set.
We can use the complex angle $\tan{\phi}=\mathcal{Y}_y/\mathcal{Y}_x$ to calculate the winding number in non-Hermitian systems, where $\phi$ can be decomposed as $\phi=\phi_r+i\phi_i$ and only $\phi_r$ affects the winding number. Thus, 
$\phi_r=\textrm{Re}\left[\tan^{-1}\left(-\frac{w\sin{k}}{v+w\cos{k}}\right)\right]$
for the system proposed here.
In particular, exceptional point in the real parts of $\mathcal{Y}_x$ and $\mathcal{Y}_y$ occur at $(0,0)$ and $\frac{1}{2\pi}\oint  \partial_k \phi_r dk$ 
%, where $k$ varies from $0$ to $2\pi$, 
takes integer multiples of $2\pi$, which implies that the winding numbers are integers. Hence we obtain four possible values for the winding number for each normal oscillation frequency: $\mu=\{-1, 0, 1, 2\}$, as shown in Figure~\ref{fig:SkinV2} by trajectories in the plane $\textrm{Re}[\mathcal{Y}_x]$, $\textrm{Re}[\mathcal{Y}_y]$ around the exceptional point. 

Physical meanings for $\mu=0$ (trivial) and $\mu=1$ (topological) in a non-Hermitian SSH model are described elsewhere~\cite{Yin2018GeometricalSystems,Lee2019AnatomySystems,Song2019Non-HermitianSystems}, we briefly discuss winding numbers $\mu=-1$ and $\mu=2$ in our model.
The winding number in a 1D topological system can be related to the quantized Zak phase. Therefore, a negative winding number ($\mu=-1$) represents a skew polarization~\cite{Mondragon-Shem2014}, i.e., occupied bands are negatively polarized, and hence the admittance matrix winds once in a clockwise direction (considering positive as meaning counterclockwise), as shown in Figure~\ref{fig:SkinV2}(c). On the other hand, winding number $\mu=2$ is related to a topological phase in the SSH model with non-negligible longer range hoppings~\cite{Perez-Gonzalez2019InterplaySystems} that connects separated sites in a sublattice, preserving chiral symmetry and allowing edge states.

%%%%%%%%%%%%

We have shown that the proposed circuit model's admittance matrix $\mathcal{Y}(k)$ adequately represents the non-Hermitian chiral-symmetric SSH system, where the quantum hopping properties are simulated in the circuit model by propagating electromagnetic modes.
Numerical simulations verified that (\ref{eq:omega}) generates integer winding numbers, which are the characteristic of an extended non-Hermitian SSH model for certain parameters (see Table~\ref{tab:table2}). 
This implies the bulk-edge correspondence, which is manifested by means of topologically protected edge states~\cite{Imura2019GeneralizedSystems}.
Indeed, the normalized eigenfunction magnitude tends to accumulate on the edge for the simulated system in Figure~\ref{fig:SkinV2} in a topological system ($\mu=1$ or $\mu=-1$), whereas no such topologically protected edge mode appears in the ordinary system ($\mu=0$). In particular, the magnitude decays slowly into the bulk when $\mu=2$~\cite{Perez-Gonzalez2018SSHDisorder}. Thus, bulk-edge correspondence relates the system winding number to the number of edge protected states, such that the number of states at each edge is equal to $|\mu|$~\cite{Maffei2018TopologicalDynamics, Imura2019GeneralizedSystems}.

\subsection{Non-Hermitian topological skin effect}

The non-Hermitian skin effect manifests itself as relocalization of bulk eigenstates towards the boundaries in certain non-Hermitian lattice systems with open edges, e.g., when the quantum hopping amplitudes bear nonreciprocity along opposite directions~\cite{MartinezAlvarez2018TopologicalSystems}. Such a phenomenon intrinsically originates from a novel non-Hermitian topology in the complex energy plane~\cite{Okuma2020TopologicalEffects,Zhang2020CorrespondenceSystems}
\begin{equation}\label{eq: winding_num}
   w(E_0) =\frac{1}{2\pi i}\int_0^{2\pi} d k \frac{d}{d k} \log \det[\mathcal{Y}(k)-E_0],
\end{equation}
where $k$ is the momentum and $E_0$ is any point on the complex energy plane. The non-Hermitian skin effect is present if and only if there exists $E_0$ on the complex energy plane such that $w(E_0)\neq0$. Because the trajectory of $\det{[\mathcal{Y}(k)-E_0]}$ in the complex plane as $k$ varies from $0$ to $2\pi$ forms a closed curve, $w(E_0)\neq0$ happens only
when the trajectory does not collapse into a retracing arc or line without interior, i.e., the closed curve has a finite interior area.
In our system, considering the non-Hermitian admittance matrix $\mathcal{Y}$'s chiral symmetry, it is convenient to analyze the trajectory of $\det{[\mathcal{Y}(k)]}$ in the complex plane.

Figure~\ref{fig:SkinV2} clearly shows the skin effect for some natural frequency solutions in the system with 300 sites and the other parameters as shown Table~\ref{tab:table2}, row 4. The absence or presence of skin effect clearly corroborates such a theoretical prediction. Note that the skin states, even if localized towards the edges, are distinguishable from the edge states. 
To this end, in Figure 6(e) we slightly alter the effective hoppings $v,w$ by 5\% at the three sites at the chain center in comparison to the case of Figure 6(b). While the Bloch and especially the blue skin states are altered, the red edge topological state remains stable.
This is because the edge states, thanks to the bulk-edge correspondence topological protection, are stable against small perturbations. On the other hand, while the existence of skin effect has a distinct topological origin, the individual skin state is not robust against such perturbations.

\section{Summary and discussion}\label{sec:conclu}
We proposed an electrical circuit model with R, L, and C elements that corresponds with a non-Hermitian SSH-like quantum model with complex hoppings and is suitable for both periodic and open boundary conditions. The system accurately exhibits various interesting topological properties, such as bulk-edge correspondence, integer winding numbers, and non-Hermitian skin effect. 
We also provide in the Appendix a realistic circuit simulation of detection in correspondence to Figure~\ref{fig:SkinV2}, where all the proposed non-Hermitian topological features can be seen.
Therefore, the current study on topolectric circuits exemplifies a promising route that relies on the correspondence between a readily available electrical circuit and topological phenomena modeling in quantum systems.

One advantage for the proposed model is that it separates intra and inter hoppings compared with previous models, offering great versatility and possibly new physical properties. The proposed model is based on natural oscillation frequencies depending on the circuit parameters values and provides a tool to analyze novel topological behavior in a non-Hermitian model.

Effectively, we are working on a non-Hermitian system with complex hopping amplitudes that converges to a Hermitian system for vanishing resistance.
Complex hopping is very useful to help understand different physical phenomena in non-Hermitian systems~\cite{Pocock2019Bulk-edgeChain, Pickup2020SyntheticPhases}. The proposed circuit provides a useful tool to analyze such systems and will help develop more interesting circuits to explore non-Hermitian phenomena.

%%%%%%%%%%%%%%%%%%%%%%%%%%%%%%%%%%

%%%%%%%%%%%%%%%%%%%%%%%%%%%%%%%%%%%%%%%%%%%%%%%%%%%%%%%
%%% Acknowledgements. ??§Ý
%%%%%%%%%%%%%%%%%%%%%%%%%%%%%%%%%%%%%%%%%%%%%%%%%%%%%%%
\section*{Acknowledgements}
D.-A.G and J.M acknowledge partial support from the Universidad de Antioquia, Colombia under initiative (CODI No.ES84180154) \textit{Estrategia de sostenibilidad del Grupo de F\'{\i}sica At\'omica y Molecular}, and projects (CODI No.251594 and No.2019-24770). X.-X.Z thanks the support from the Riken Special Postdoctoral Researcher Program and the Max Planck-UBC-UTokyo Center for Quantum Materials.

%%%%%%%%%%%%%%%%%%%%%%%%%%%%%%%%%%%%%%%%%%%%%%%%%%%%%%%
%%% Conflict of interest. ????????????
%%%%%%%%%%%%%%%%%%%%%%%%%%%%%%%%%%%%%%%%%%%%%%%%%%%%%%%
%\InterestConflict{The authors declare that they have no conflict of interest.}

\bibliographystyle{unsrt}
\bibliography{references}

%%%%%%%%%%%%%%%%%%%%%%%%%%%%%%%%%%%%%%%%%%%%%%%%%%%%%%%
%%% Appendix sections. ??????, ????
%%%%%%%%%%%%%%%%%%%%%%%%%%%%%%%%%%%%%%%%%%%%%%%%%%%%%%%
\begin{appendix}

\section{Circuit simulation}
\begin{figure}[h!]
\centering
\includegraphics[scale=0.48]{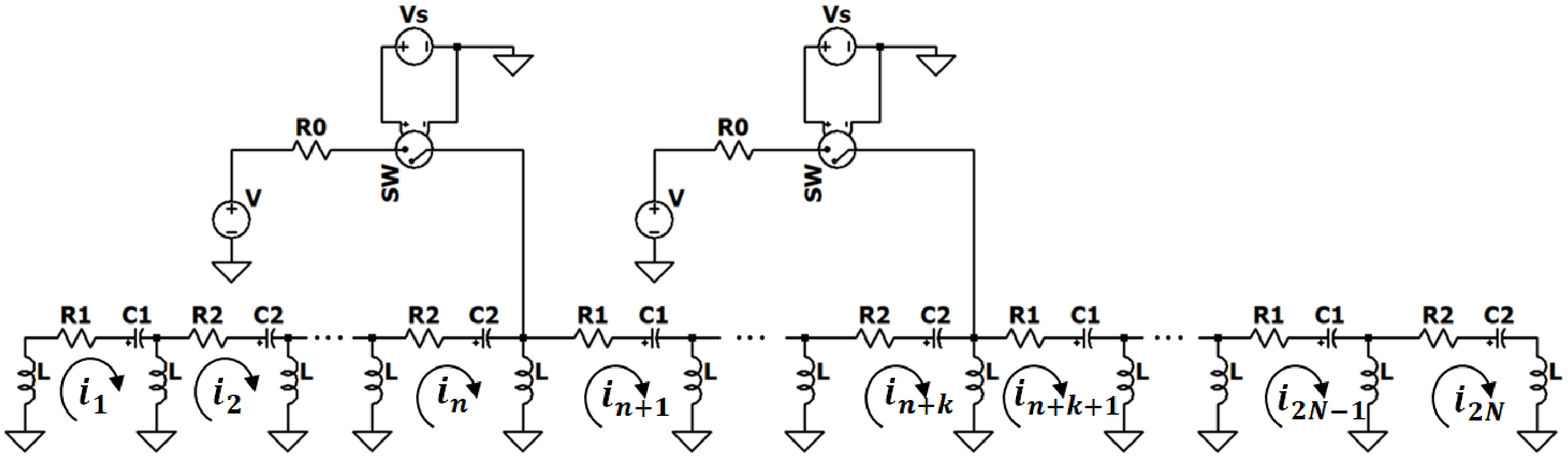}
\caption{Circuit diagram to simulate the non-Hermitian topological model. Both the two voltage sources $(V)$ oscillate at one of the natural frequencies and they are located at 1/3 and 2/3 positions of the whole circuit chain. Once the circuit is oscillating at such frequency, the switch (SW) opens and after the transient period, the current response is observed from the energy stored in the different capacitors and inductors of the circuit.
}
\label{fig:Experim}
\end{figure}

\begin{figure*}[b]
\centering
\includegraphics[scale=0.58]{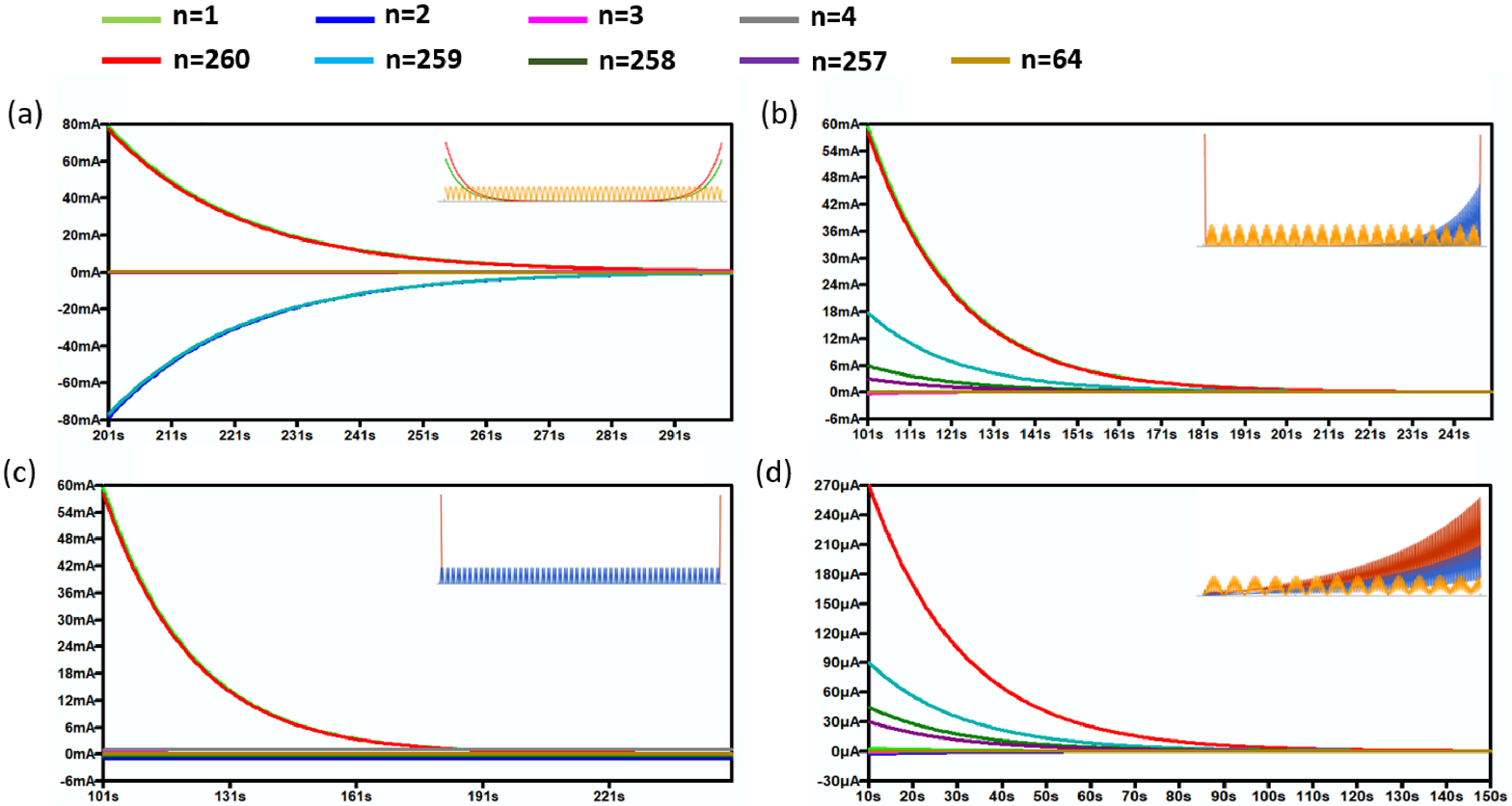}
\caption{Simulation for the RLC parameters of row 4 of Table \ref{tab:table2} for $N = 260$. Current through the grounded inductors at each node $n$ are measured after the initial transient process due to opening of the switches. In the upper right corner of each figure we show the shape of the magnitude of the eigenvectors in accordance with Figure \ref{fig:SkinV2}. Here, (a,b,c,d) respectively correspond to the excitation and detection of the situations in Figure \ref{fig:SkinV2}(a,b,c,d).
}
\label{fig:result}
\end{figure*} 

In figure \ref{fig:Experim} we show the simulated circuit diagram. The voltage sources $V$ are set to the natural frequency of oscillation and the process of charging/discharging of capacitors and inductors begins throughout the entire network. 
Once we open the switches, the circuit is now oscillating at the natural frequency of oscillation $\omega=\omega_R-i\omega_I$ given by Equation (\ref{eq:omega}).
After the transient process due to the opening of the source switch, begins the gradual dissipation of the energy stored in capacitors and inductors due to the resistors.

This causes the voltage and current signal to decay over time in the generic form $Ae^{-\omega_I t}\cos{(\omega_R t+\phi)}$ of a damped oscillator. As $\omega_I$ is physically guaranteed to be positive due to the dissipative nature, the characteristic relaxation time scale is given by $1/\omega_I$.
The wide observation window extends until the circuit signals is reduced to lie outside of the detection reach. Simply extracting the ground current signals enables us to observe the non-Hermitian topological edge states and skin phenomenon.

We show the simulation results from LTSpice with parameters of row 4 of table \ref{tab:table2} that we also use in Figure \ref{fig:SkinV2} in one-to-one correspondence. In Figures \ref{fig:result}(a) and (c), we observe that only currents around the chain edges are significantly excited although the sources are located very far away. This clearly shows the presence of the non-Hermitian topological edge states at both ends. Note that even the different spatial decay rate of at different topological phases can be clearly observed: it is almost only the two end nodes are excited in Figure \ref{fig:result}(c) in good match with the theoretical prediction.
In Figures \ref{fig:result}(c) and (d), the skin states appear and manifest themselves as the signal around the chain ends. Note that they decay in space more slowly than the edge states and hence the signals from more boundary nodes than Figures \ref{fig:result}(a) and (c). It is crucial to note that only the right boundary exhibits the signal in Figure \ref{fig:result}(d), which clearly proves the appearance of skin states at only one end as theoretically predicted. In Figure \ref{fig:result}(c), there is as well a sharp green signal from the left boundary, which is exactly due to the coexistence of edge modes and skin states.

\end{appendix}

%\end{multicols}
\end{document}